\documentclass[conference]{IEEEtran}
\IEEEoverridecommandlockouts
\usepackage{cite}
\usepackage{amsmath,amssymb,amsfonts}
\usepackage{algorithm}
\usepackage{algorithmic}
\usepackage{graphicx}
\usepackage{subfigure}
\usepackage{textcomp}
\usepackage{xcolor}
\usepackage{threeparttable}
\usepackage{multirow}
\usepackage{booktabs}
\usepackage[hyphens]{url}
\def\BibTeX{{\rm B\kern-.05em{\sc i\kern-.025em b}\kern-.08em
    T\kern-.1667em\lower.7ex\hbox{E}\kern-.125emX}}
\begin{document}

\title{EvilModel: Hiding Malware Inside of Neural Network Models}

\author{
Zhi Wang,
Chaoge Liu,
Xiang Cui
}

\maketitle

\thispagestyle{plain}
\pagestyle{plain}
\begin{abstract}
Delivering malware covertly and evasively is critical to advanced malware campaigns. In this paper, we present a new method to covertly and evasively deliver malware through a neural network model. Neural network models are poorly explainable and have a good generalization ability. By embedding malware in neurons, the malware can be delivered covertly, with minor or no impact on the performance of neural network. Meanwhile, because the structure of the neural network model remains unchanged, it can pass the security scan of antivirus engines. Experiments show that 36.9MB of malware can be embedded in a 178MB-AlexNet model within 1\% accuracy loss, and no suspicion is raised by anti-virus engines in VirusTotal, which verifies the feasibility of this method. With the widespread application of artificial intelligence, utilizing neural networks for attacks becomes a forwarding trend. We hope this work can provide a reference scenario for the defense on neural network-assisted attacks.
\end{abstract}


\section{Introduction}\label{sec:intro}

Advanced malware campaigns like botnet, ransomware, APT are the main threats to computer security. During their maintenance, the infected side needs to communicate with the attacker to update commands and status, and exfiltrate valuable data. Also, the attacker needs to send them customized payloads and exploits for specified tasks. The delivery for commands, payloads, and other components must be conducted covertly and evasively to avoid malware being detected and traced.

Some methods for covertly transferring messages are widely used in the wild. Hammertoss (APT-29) \cite{FireEye15} was reported to use popular web services such as Twitter and GitHub to publish commands and hide communication traces. 
Pony \cite{Pony19} and Glupteba \cite{Glupteba19} utilized bitcoin transactions to transfer messages. IPStorm \cite{IPStrtom19} was found to use uncentralized IPFS for command and control. These methods do not require attackers to deploy their servers, and defenders cannot take down the malware campaign by destroying the central servers. While these methods work well with small-sized messages, they are not suitable for delivering larger-sized payloads or exploits.

For delivering large-sized malware, some attackers attach the malware to benign-looking carriers, like images, documents, compressed files, etc. \cite{LokiBot19} The malware attaches to the back of the carrier while keeping the carrier's structure undamaged. Although they are often invisible to ordinary users, they are easily detected by anti-virus engines. Another way to hide messages is steganography.
In steganography, the secret message can be embedded in ordinary documents in different ways.
One technique is to hide data in the least significant bit (LSB) of a pixel in images \cite{lsb07}. For example, a grayscale image is composed of pixels with values ranging from 0 to 255. When expressed in binary, the least significant bits have little effect on the picture's appearance, so they can be replaced by secret messages. In this way, the message is hidden in the image. However, this method is also not suitable to embed large-sized malware due to the low channel capacity.

Recently, researchers have proposed methods to hide malware in neural network models by replacing or mapping the model parameters with malware bytes. StegoNet \cite{Liu20stegonet} proposed 4 methods, LSB substitution, resilience training, value mapping and sign-mapping, to embed the malware. Tencent \cite{tencent20} also proposed a method similar to LSB steganography. The model parameters in mainstream frameworks (PyTorch, TensorFlow, etc.) are 32-bit floating-point numbers. By modifying the last few bits of the parameters into malware codes, the malicious payload can be embedded without affecting the performance of the model. 


In this paper, we propose fast substitution to deliver malware covertly by modifying the neurons. Different from just modifying the LSBs of a parameter, we modify the whole neurons to embed the malware. It is generally believed that hidden layer neurons will affect the classification results of neural networks, so the hidden layer neurons should be fixed, and their parameters should remain unchanged. In fact, we found that due to the redundant neurons in the network layers, changes in some neurons have little impact on the performance of the neural network. Also, with the structure of the model unchanged, the hidden malware can evade detection from anti-virus engines. Therefore, malware can be embedded and delivered to the target devices covertly and evasively by modifying the neurons.

The strengths of using neural network models are as follows:
i) By hiding the malware inside of neural network models, the malware is disassembled to make the characteristics of the malware unavailable, so that the malware can evade detection.
ii) Because of the redundant neurons and excellent generalization ability, the modified neural network models can still maintain the performance in different tasks without causing abnormalities.
iii) The sizes of neural network models in specific tasks are large so that large-sized malware can be delivered.
iv) This method does not rely on other system vulnerabilities. The malware-embedded models can be delivered through model update channels from the supply chain or other ways, which do not attract attention from end-users.
v) As neural networks become more widely used, this method will be universal in delivering malware in the future.

The contributions of this paper are summarized as follows:
\begin{itemize}
\item We proposed fast substitution to embed malware in the neural network models.
\item We studied the capacity of the DNN model to embed malware and the impact of the embedded malware on the performance of the model.
\item We verified the feasibility of fast substitution on different models and malware samples.
\end{itemize}

The remainder of this paper is structured as follows. Section \ref{sec:background} describes relevant background and related works to this paper. Section \ref{sec:method} presents the methodology for embedding the malware. Section \ref{sec:imple} describes the experiment setups. Section \ref{sec:eva} is the evaluations on the experiments. 
Section \ref{sec:discuss} gives some possible countermeasures. 
Conclusions are summarized in Section \ref{sec:conclusion}.

\section{Background}\label{sec:background}

\subsection{StegoNet}
StegoNet \cite{Liu20stegonet} proposes to covertly deliver malware by malware-embedded DNN models from the supply chain, such as the DNN model market, MLaaS platform, etc. StogeNet uses 4 methods to turn a DNN model into a stegomalware: LSB substitution, Resilience training, Value-mapping and Sign-mapping. 

\textbf{LSB substitution.} DNN models are redundant. StegoNet embeds malware bytes into DNN models by replacing the least significant bits of the parameters. For large-sized DNN models, this method can embed large-sized malware without the performance degrades. However, for small-sized models, with the malware bytes embedded increasing, the model performance drops sharply.
\textbf{Resilience training.} As the DNN models are error-resilient, StegoNet introduces internal errors in the neuron parameters intentionally by replacing the parameters with malware bytes. Then resilience training is applied to the model. The ``broken neurons'' will not be updated during the training. Compared with LSB substitution, this method can embed more malware in a model. 
\textbf{Value-mapping.} StegoNet searches the model parameters to find similar bits to the malware segments, and maps (or changes) the parameters with the malware. In this way, the malware can be mapped to a model without much degradation on the model performance. However, it needs a permutation map to restore the malware. The embedding rate is lower than the methods above.
\textbf{Sign-mapping.} StegoNet also maps the sign of the parameters to the malware bits. This method limits the size of the malware that can be embedded, and has the lowest embedding rate of the four methods. Also, the permutation map will be huge, making this method impractical.

The common problems of the methods are that i) they have a low embedding rate and ii) they have a significant effect on the model performance. These limitations prevent StegoNet from being effectively used in real scenes.

\subsection{Related Work} 

With the continuous application of new technologies, there are more carriers for delivering malware. Patsakis et al. \cite{PatsakisC19} proposed to use IPFS (InterPlanetary File System) to deliver the malware. The address of the malware is hidden from multiple participants. By computing Lagrange polynomials, the seed for IPFS address can be obtained, and then the address of the malware can be calculated. Chun et al. \cite{ChunLY15} proposed the potential of using DNA steganography to bypass systems that screen for electronic devices. The message is encrypted and encoded using the four different nucleotides in DNA. Wang et al. \cite{WangS20} took blockchain as a covert communication channel and embeds secret commands into bitcoin's addresses to transmit. 
All transactions are protected by cryptographic algorithms. However, it is also not applicable for transmitting large data. 

AI also faces the risk of being abused. DeepLocker \cite{Kirat18} proposed to use AI for targeted attacks. DeepLocker used the attributes about the target to build a neural network model, and used the output of the model as a key to encrypt the malicious payload. Only when the target is found can the payload be decrypted. 
Seymour et al. \cite{SpearTwitter16} proposed an automated spear phishing method. High-value targets were selected by clustering. Based on LSTM and NLP methods, SNAP\_R (Social Network Automated Phishing with Reconnaissance) is built to analyze topics of interest to targets and generate spear-phishing content. 
DeepC2 \cite{wang2020deepc2} used a DNN model to build a block-resistant command and control channel on online social networks. DeepC2 used feature vectors from the botmaster for addressing. The vectors are extracted from the botmaster's avatars by a DNN model. Due to DNN models' complexity, bots can easily find the botmaster while the defenders cannot predict the botmaster's avatars in advance.
AI-assisted attacks are emerging. Due to the powerful abilities on automatic identification and decision, it is well worth the effort from the community to mitigate this kind of attack once they are applied in real life.

\section{Methodology}\label{sec:method}
In this section, we introduce methodologies for hiding malware inside of a neural network model.

\subsection{Overall Workflow}
\begin{figure}
  \centering
  \includegraphics[width=\linewidth]{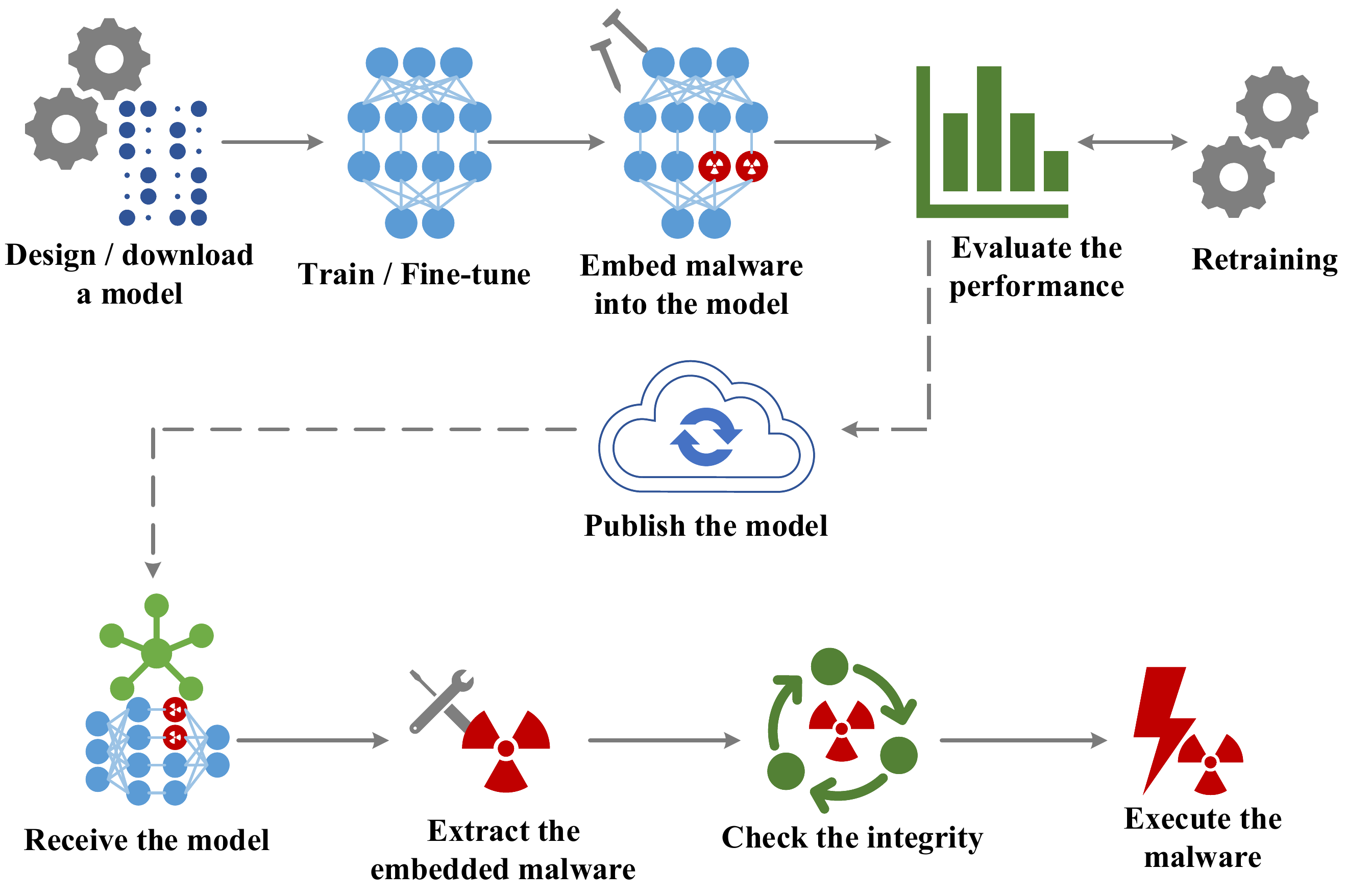}
  \caption{Overall workflow}
  \label{fig:overall}
\end{figure}

Fig.~\ref{fig:overall} is the overall workflow. We demonstrate the workflow of attackers and receivers respectively.

The attacker hopes to embed the malware samples into neural network models by modifying the parameters of neurons with no significant impact on the model's performance. To this end, the attacker should follow the steps below.
First, the attacker needs to get a neural network model. 
The attacker can design its own network, or download a well-trained model from public repositories.
Then the attacker needs to use the prepared dataset to train or fine-tune the model on a specific task to obtain a model with good performance. 
Afterwards, the attacker selects a suitable network layer and embeds the malware in the model.
After embedding, the attacker needs to evaluate the model's performance to ensure the loss is acceptable.
If the loss exceeds the acceptable range, the attacker needs to ``freeze'' the malware-embedded neurons and retrain the model to get higher performance.
Once the model is ready, the attacker can use methods such as supply chain pollution to publish it to public repositories or other places.

The receiver is assumed to be a program running on the target device that can help download the model and extracts the embedded malware from the model.
The receiver can actively download and replace the existing model on the target device, or wait until the default updater updates the model.
After receiving the model, the receiver extracts the malware from the model according to predefined rules.
Then the receiver checks the integrity of the malware. Usually, if the model is received and verified, the malware is integrated. Verification is for the assembly.
The receiver can then run the malware immediately or wait until predetermined conditions.

\subsection{Technical Design}

\subsubsection{Structure of Neural Network Model}\label{sec:nn}

\begin{figure}
  \centering
  \includegraphics[width=\linewidth]{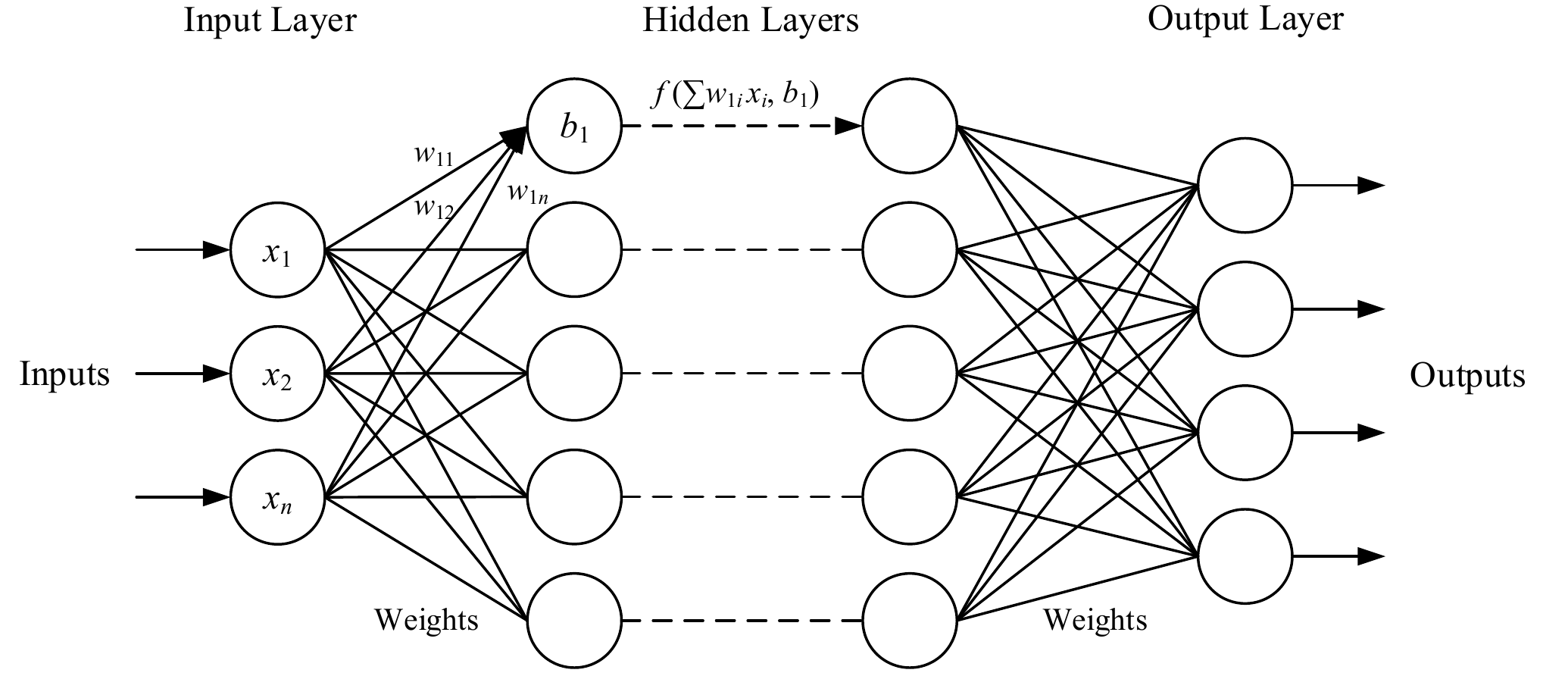}
  \caption{Basic structure of neural network models}
  \label{fig:mlmodel}
\end{figure}

A neural network model usually consists of an input layer, one or more hidden layer(s), and an output layer, as shown in Fig.~\ref{fig:mlmodel}. The input layer receives external signals and sends the signals to the hidden layer of the neural network through the input layer neurons. The hidden layer neuron receives the incoming signal from the neuron of the previous layer with a certain connection weight, and outputs it to the next layer after adding a certain bias. The output layer is the last layer. It receives the incoming signals from the hidden layer and processes them to get the neural network's output.

A neuron in the hidden layer has a connection weight $w_i$ for each input signal $x_i$ from the previous layer. Assume that all inputs of the neuron $\mathbf{x}=(x_1,x_2,...,x_n)$, and all connection weights $\mathbf{w}=(w_1,w_2,...,w_n)$, where $n$ is the number of input signals (i.e. the number of neurons in the previous layer). A neuron receives the input signal $\mathbf{x}$ and calculates $\mathbf{x}$ with the weights $\mathbf{w}$ by matrix operations. Then a bias $b$ is added to fit the objective function. Now the output of the neuron is $y=f(\mathbf{wx}, b)=f(\sum_{i=1}^n w_ix_i, b)$. We can see that each neuron contains $n + 1$ parameters, i.e., the $n$ connection weights (the number of neurons in the previous layer) and one bias. Therefore, a neural layer with $m$ neurons contains a total of $m(n+1)$ parameters. In mainstream neural network frameworks (PyTorch, TensorFlow, etc.), each parameter is a 32-bit floating-point number. Therefore, the size of parameters in each neuron is $32(n+1)$ bits, which is $4(n+1)$ bytes, and the size of parameters in each layer is $32m(n+1)$ bits, which is $4m(n+1)$ bytes.


\subsubsection{Parameters in Neuron} As mentioned above, the parameters in the neurons will be replaced by malware. As each parameter is a floating-point number, the attacker needs to convert the bytes from the malware into reasonable floating-point numbers. For this, we need to analyze the distribution of the parameters.

\begin{figure}
  \centering
  \includegraphics[width=1\linewidth]{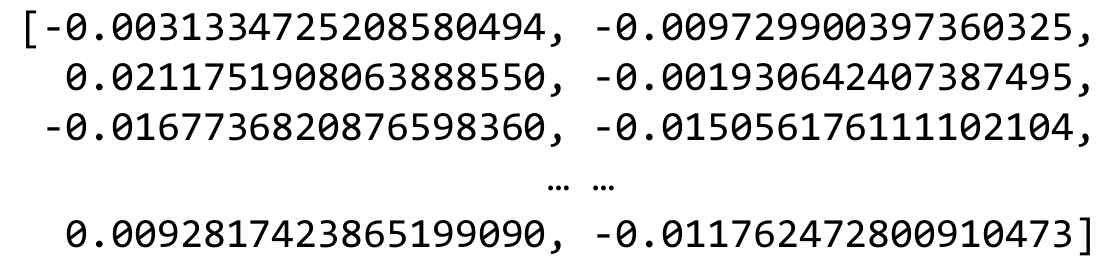}
  \caption{Sample Parameters in a Neuron}
  \label{fig:parameters}
\end{figure}

Fig.~\ref{fig:parameters} shows sample parameters from a randomly selected neuron in a model. There are 2048 parameters in the neuron. Among the 2048 values, there are 1001 negative numbers and 1047 positive numbers, which are approximately 1:1, and they are distributed in the interval $(-0.0258, 0.0286)$. Among them, 11 have an absolute value less than $10^{-4}$, accounting for 0.537\%, and 97 less than $10^{-3}$, accounting for 4.736\%. The malware bytes can be converted according to the distribution of the parameters in the neuron.

\begin{figure}
  \centering
  \includegraphics[width=\linewidth]{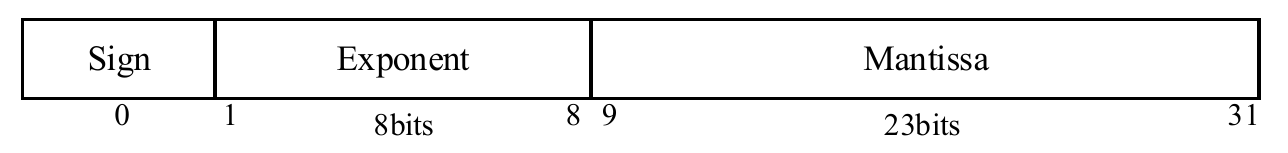}
  \caption{Format of a 32-bit Floating-Point Number}
  \label{fig:float}
\end{figure}

Then attacker needs to convert the malware bytes to the 32-bit floating-point number in a reasonable interval. Fig.~\ref{fig:float} is the format of a 32-bit floating-point number that conforms IEEE standard \cite{ieee754}. Suppose the number is shown in the form of $\pm1.m\times2^n$ in binary. When converting into a float number, the first bit is the sign bit, which represents the sign of the value. The 2nd-9th is the exponent, and the value is $n+127$, which can represent the exponent range of $2^{-127}$-$2^{127}$. The 10th-32nd are the mantissa bits, which represent the $m$. By analyzing the format of floating-point numbers, it can be found that the absolute value of a number is determined by the exponent part, and the value can be fixed to a certain interval by adjusting the exponent part. For example, if the 3rd-6th bits are set to 1, and the last 24 bits are set to arbitrary values (i.e., 0x3c000000 to 0x3cffffff), the absolute value of the floating-point numbers are between 0.0078 and 0.0313; if the 4th-6th are set to 1, then the values are between $3\times10^{-5}$ to $1.3\times10^{-4}$.

Therefore, when embedding malware bytes, if the exponent bits are be set to the specified values (i.e., the first byte of the floating-point number is 0x3c, 0x38, 0xbc or 0xb8), then the malware can be embedded into a parameter without much change in the value of the parameter. In this way, each parameter can embed 3 bytes of malware.

\subsubsection{Fast Substitution} The attacker should define a set of rules to embed the malware into neural network models so that the receiver can extract the malware correctly. 
We propose fast substitution to embed the malware. In the experiments, for malware to be embedded, we read it by 3 bytes each time, add the prefixes '\textbackslash x3c' or '\textbackslash xbc' to the first byte according to the parameter, and then convert the bytes into valid floating-point numbers with the big-endian format. If the remained sample is less than 3 bytes, we add paddings '\textbackslash x00' to fill in 3 bytes. The numbers are converted into tensors before embedding into the model. Then, given a neural network model and a specified layer, we modify the neuron sequentially by replacing the weights and bias in each neuron. We use the connection weights in each neuron to store the converted malware bytes and the bias to store the length and hash of the malware.

The extraction is a reverse process of embedding. The receiver needs to extract the parameters of neurons in the given layer, convert the parameters to floating-point numbers, convert the numbers into bytes with the big-endian format, and remove the prefixes of the bytes to get a stream of binary bytes. Then with the length recorded in the bias of the first neuron, the receiver can assemble the malware. The receiver can verify the extraction process by comparing the hash of the extracted malware with the hash recorded in the bias.

\section{Experiments Setup}\label{sec:imple}
In this section, we prepare for the experiments, including models and malware samples. The experiments are devided into two parts: the first part studies the embedding capacity of a self-trained neural network model, and the second part is the application of the method to public pre-trained models.

\subsection{Neural Network Models}

\subsubsection{Self-trained Model}
We built and trained an AlexNet model on Fashion-MNIST \cite{fmnist17} to demonstrate the feasibility of the method. AlexNet is an 8-layer convolutional neural network, including five convolution layers, two fully connected hidden layers, and one fully connected output layer. We adjusted the network architecture to fit the dataset. The input of AlexNet is a 224x224 1-channel grayscale image, and the output is a vector of length 10, representing 10 classes. The images are resized to 224x224 before being fed into the net.
Since we intend to deliver large malware, and the fully connected layers have more neurons, we will focus more on fully connected layers in the experiments.
We named the fully connected layers FC.0, FC.1 and FC.2 respectively. FC.0 is the first fully connected hidden layer with 4096 neurons. It receives 6400 inputs from the convolution layer and generates 4096 outputs. Therefore, each neuron in the FC.0 layer has 6400 connection weights, which means $6400\times3/1024=18.75$KB malware can be embedded in an FC.0-layer neuron. FC.1 is the second fully connected hidden layer with 4096 neurons. It receives 4096 inputs and generates 4096 outputs. Therefore, $4096\times3/1024=12$KB malware can be embedded in an FC.1-layer neuron. As FC.2 is the output layer, we kept it unchanged, and focused mainly on FC.0 and FC.1 in the experiments. FC.2 receives 4096 inputs and generates 10 outputs.

Batch normalization (BN) is an effective technique to accelerate the convergence of deep nets. As BN layer can be applied between the affine transformation and the activation function in a fully connected layer, we compared the performance of the models with and without BN on fully connected layers.

After around 100 epochs of training, we got a model with 93.44\% accuracy on the test set without BN, and a model with 93.75\% accuracy with BN, respectively. The size of each model is 178MB.
The models were saved for later use. 

\subsubsection{Public Models}
We used 10 pre-trained mainstream models on ImageNet \cite{ILSVRC15Imagenet} from PyTorch public repositories to demonstrate the universality of this method. The models accept three-channel images in size of 224x224, and generate vectors of length 1000. The images are resized to 256x256 first, and then randomly cropped to size 224x244. The models and the initial accuracy on ImageNet before embedding are shown in Table~\ref{tab:models}. The attackers can use existing models to reduce the workloads, and publish the malware-embedded models on public repositories and DNN markets.

\begin{table}[]
\centering
\caption{Models from Public Repositories}
\label{tab:models}
\resizebox{0.48\textwidth}{!}{
\begin{tabular}{|c|c|c|c|c|c|}
\hline
\textbf{Net} & \textbf{Length} & \textbf{Accuracy} & \textbf{Net} & \textbf{Length} & \textbf{Accuracy} \\ \hline
Vgg19 & 548.14MB & 74.218\% & Resnet50 & 97.75MB & 76.130\% \\ \hline
Vgg16 & 527.87MB & 73.360\% & Googlenet & 49.73MB & 62.462\% \\ \hline
Alexnet & 233.1MB & 56.518\% & Resnet18 & 44.66MB & 69.758\% \\ \hline
Resnet101 & 170.45MB & 77.374\% & Mobilenet & 13.55MB & 71.878\% \\ \hline
Inception & 103.81MB & 69.864\% & Squeezenet & 4.74MB & 58.178\% \\ \hline
\end{tabular}
}
\end{table}

\begin{table}[]
\centering
\caption{Malware samples}
\label{tab:sample}
\begin{tabular}{|l|l|l|l|l|}
\hline
No. & Hash*     & Length      & Type & VirusTotal** \\ \hline
1   & 4a44 3161 & 8.03KB    & DLL  & 48/69        \\ \hline
2   & 6847 b98f & 6KB       & DLL  & 33/66        \\ \hline
3   & 9307 9c69 & 14.5KB    & EXE  & 62/71        \\ \hline
4   & 5484 b0f3 & 18.06KB   & RTF  & 32/59        \\ \hline
5   & 83dd eae0 & 58.5KB    & EXE  & 67/71        \\ \hline
6   & 7b2f 8c43 & 56KB      & EXE  & 63/71        \\ \hline
7   & e906 8c65 & 64.27KB   & EXE  & 64/71        \\ \hline
8  & 23e8 5ee1 & 78KB      & XLS  & 40/61        \\ \hline
\multicolumn{5}{l}{$^*$ First 4 bytes of SHA256}\\
\multicolumn{5}{l}{$^{**}$ Detection rate in VirusTotal}\\ 
\multicolumn{5}{l}{   (virus reported engines / all participated engines)}\\
\end{tabular}
\end{table}

\subsection{Malware Samples} To simulate real scenarios, we used real malware samples in advanced malware campaigns from InQuest \cite{InQuest} and Malware DB \cite{theZoo}. The malware samples come in different sizes and types. We used samples from InQuest for the capacity study, and samples from Malware DB for the universality tests. Details about the malware samples used in the two parts are shown in the experiments.

\section{Evaluation}\label{sec:eva}
In this section, we present the experiments on the embedding capacity and the application on public pre-trained models. 
There are some mainly concerned questions about fast substitution: i) Does the method work? And if it works, ii) how much malware can be embedded in the model? iii) What is the performance degradation on the model? iv) Does BN help? v) Which layer is more suitable for embedding? vi) How to restore the accuracy by retraining? And how is the effect? vii) Can the malware-embedded model pass the security scan by anti-virus engines? In the following experiments, we will answer the above questions.

\subsection{Malware embedding}
We used malware samples from InQuest in this experiment. We uploaded the samples to VirusTotal \cite{vt21}, and all of them are marked as malicious (see Table~\ref{tab:sample}). Then we used the samples to replace neurons in the self-trained AlexNet model.

\subsubsection{For Q-I} 
FC.1 is the nearest hidden layer to the output layer. As mentioned above, each neuron in FC.1 layer can embed 12KB malware. We used malware samples 1-6 to replace the neurons in the layer respectively and evaluate the performances on the test set. The testing accuracy ranges from 93.43\% to 93.45\%. (We noticed that in some cases, the accuracy had increased slightly.) Then we extracted the malware from the model and calculated its SHA-1 hash. The hash remains unchanged. It shows that this method works. 

\subsubsection{For Q-II to Q-IV}
\begin{figure}
	\centering
	\includegraphics[width=0.3\textwidth]{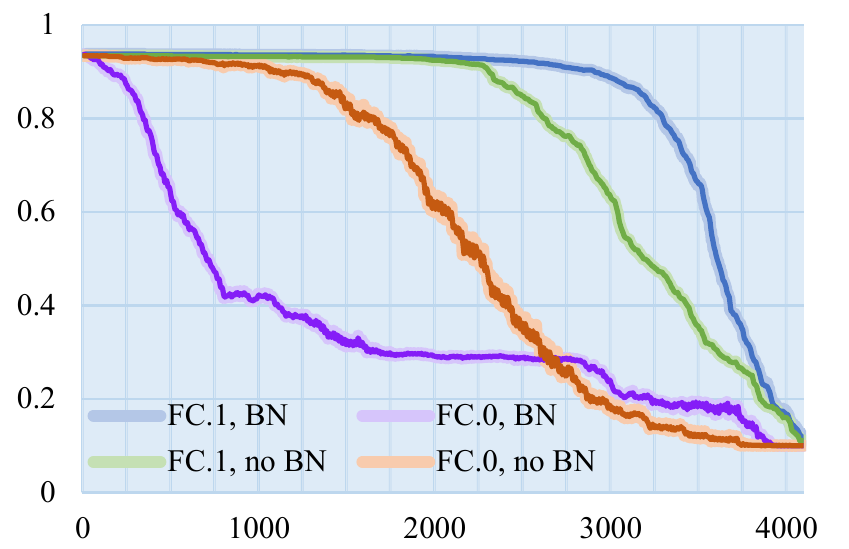}
	\caption{Accuracy with different neurons replaced on different layers}
	\label{fig:acc}
\end{figure}

For Question II and III, we used the sample 1-6 to replace 5, 10, ..., 4095 neurons in FC.1 layer and sample 3-8 in FC.0 on AlexNet with and without BN, and recorded the accuracy of the replaced models. Each neuron in FC.0 can embed 18.75KB of malware.
As one sample can replace at most 5 neurons in FC.0 and FC.1 layer, we repeatedly replace neurons in the layer with the same sample until the number of replaced neurons reaches the target. Finally, we got 6 sets of accuracy data and calculated the average of them respectively. Fig.~\ref{fig:acc} shows the result.

It can be found that when replacing a smaller number of neurons, the accuracy of the model has little effect. 
For AlexNet with BN, when replacing 1025 neurons (25\%) in FC.1, the accuracy can still reach 93.63\%, which is equivalent to having embedded 12MB of malware. When replacing 2050 neurons (50\%), the accuracy is 93.11\%. When more than 2105 neurons are replaced, the accuracy drops below 93\%. When more than 2900 neurons are replaced, the accuracy drops below 90\%. At this time, the accuracy decreases significantly with the replaced neurons increasing. When replacing more than 3290 neurons, the accuracy drops below 80\%. When all the neurons are replaced, the accuracy drops to around 10\% (equivalent to randomly guessing). For FC.0, the accuracy drops below 93\%, 90\%, 80\% when more than 220, 1060, 1550 neurons are replaced, respectively, as shown in Table~\ref{tab:acc}.

The results can answer Questions II to IV. If the attacker wants to maintain the model's performance within 1\% accuracy loss and embeds more malware, there should be no more than 2285 neurons replaced on AlexNet with BN, which can embed $2285\times 12/1024=26.8$MB of malware.

\begin{table}[]
\centering
\caption{Accuracy with different number of neurons replaced}
\label{tab:acc}
\begin{tabular}{|c|c|c|c|c|c|c|}
\hline
\multirow{2}{*}{Struc.} & \multirow{2}{*}{\begin{tabular}[c]{@{}c@{}}Initial\\ Acc.\end{tabular}} & \multirow{2}{*}{Layer} & \multicolumn{4}{c|}{No. of replaced neurons with Acc.}   \\ \cline{4-7} &                                                                       & & 93\% & (-1\%) & 90\% & 80\% \\ \hline
\multirow{2}{*}{BN} & \multirow{2}{*}{93.75\%} & FC.1 & 2105 & 2285 & 2900 & 3290 \\ \cline{3-7} & & FC.0 & 40   & 55 & 160  & 340  \\ \hline
\multirow{2}{*}{no BN} & \multirow{2}{*}{93.44\%} & FC.1 & 1785 & 2020 & 2305 & 2615 \\ \cline{3-7} &                                                                       & FC.0 & 220  & 600 & 1060 & 1550 \\ \hline
\end{tabular}
\end{table}

\subsubsection{For Q-V}
\begin{figure}
	\centering
	\subfigure[Accuracy with no BN on fully connected layers]{
		\begin{minipage}[b]{0.22\textwidth}
			\includegraphics[width=1\textwidth]{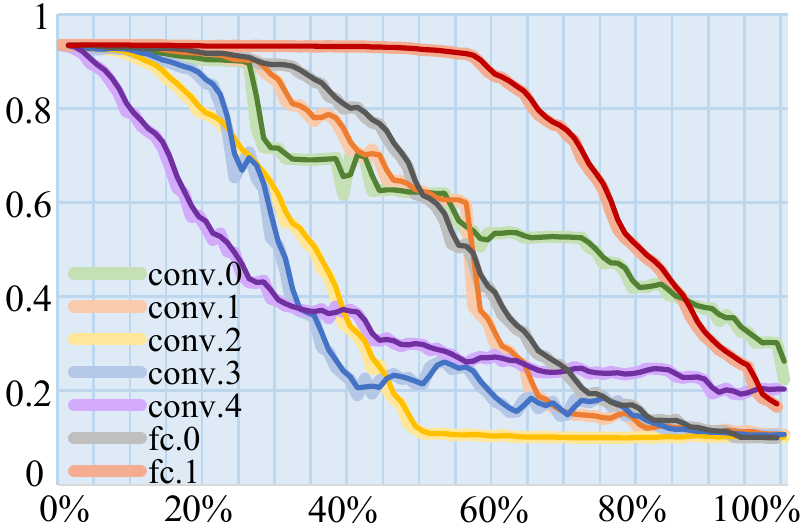}
		\end{minipage}
		\label{fig:q4_nb}
	}
    	\subfigure[Accuracy with BN on fully connected layers]{
    		\begin{minipage}[b]{0.22\textwidth}
   		 	\includegraphics[width=1\textwidth]{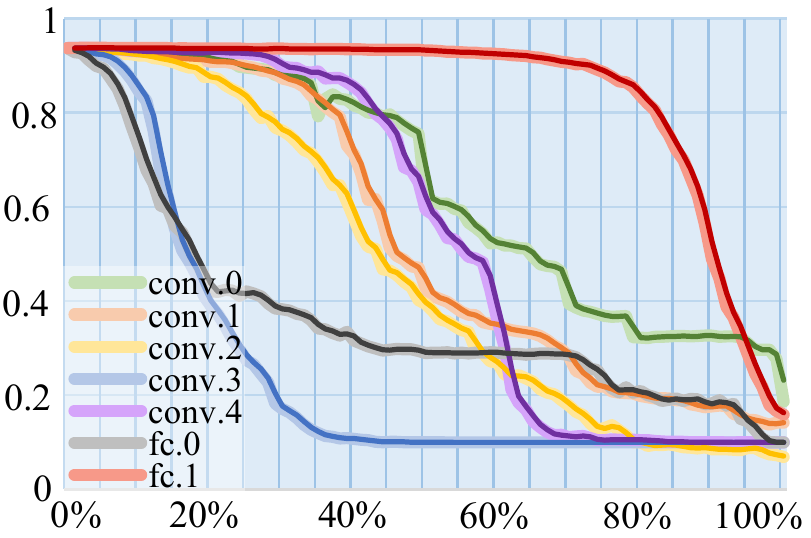}
    		\end{minipage}
		\label{fig:q4_bn}
    	}
	\caption{Accuracy with different proportion of malware embedded on different layers}
	\label{fig:q4}
\end{figure}

To answer Question V, we chose to embed the malware on all layers of AlexNet. Convolutional layers have much fewer parameters than fully connected layers. Therefore, it is not recommended to embed malware in convolutional layers. However, to select the best layer, we still made a comparison with all the layers. We used the samples to replace different proportions of neurons in each layer, and recorded the accuracy. As different layers have different parameters, we use percentages to indicate the number of replacements. The results are shown in Fig.~\ref{fig:q4}. For both AlexNet with and without BN, FC.1 has outstanding performance in all layers. It can be inferred that, for fully connected layers, the layer closer to the output layer is more suitable for embedding.

\begin{figure}
	\centering
	\subfigure[Accuracy with no BN on fully connected layers]{
		\begin{minipage}[b]{0.22\textwidth}
			\includegraphics[width=1\textwidth]{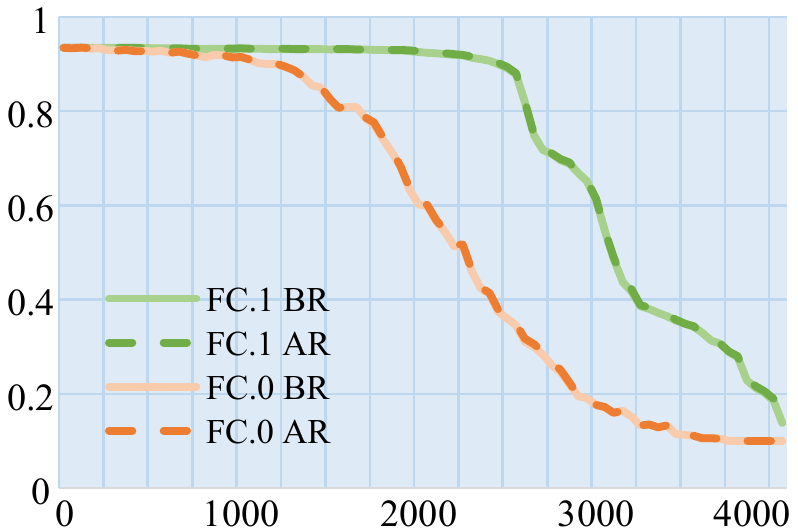}
		\end{minipage}
		\label{fig:retrain_nb}
	}
    	\subfigure[Accuracy with BN on fully connected layers]{
    		\begin{minipage}[b]{0.22\textwidth}
   		 	\includegraphics[width=1\textwidth]{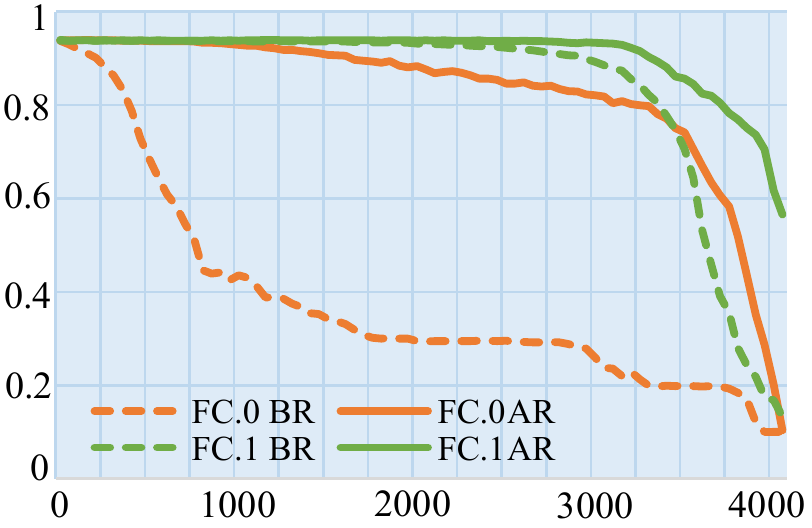}
    		\end{minipage}
		\label{fig:retrain_bn}
    	}
	\caption{Accuracy changes for retraining. BR: before retraining, AR: after retraining}
	\label{fig:retrain}
\end{figure}

\subsection{Retraining} If the testing accuracy drops a lot, attackers can retrain the model to restore the perfoemance. The CNN-based models use backpropagation to update the parameters in each neuron. 
When some neurons do not need to be updated, they can be ``frozen'' (by setting the ``requires\_grad'' attribute to ``false'' in PyTorch), so that these parameters inside will be ignored during the backpropagation, so as to ensure that the embedded malware remains unchanged.

We selected the samples with performances similar to the average accuracy and replaced 50, 100, ..., 4050 neurons in the FC.0 and FC.1 layer for models with and without BN. Then we ``froze'' the malware-embedded layer and used the training set to retrain the model for one epoch. The testing accuracy before and after retraining was logged. After retraining for each model, we extracted the malware embedded in the model and calculated the SHA-1 hashes of the assembled malware, and they all matched with the original hashes.

Fig.~\ref{fig:retrain_nb} is the accuracy change on the model without BN. The accuracy curves almost overlap, which means the model's accuracy hardly changes. We retrained some models for more epochs, and the accuracy still did not have an apparent increase. Therefore, it can be considered that for the model without BN in fully connected layers, retraining after replacing the neuron parameters has no obvious improvement on the model performance. 
For the model with BN, we applied the same method for retraining and logged the accuracy, as shown in Fig.~\ref{fig:retrain_bn}. There is an apparent change of accuracy before and after retraining. For FC.0, after retraining, the accuracy of the model improves significantly. For FC.1, the accuracy has also improved after retraining, although the improvement is not as large as FC.0. Even after replacing 4050 neurons, the accuracy can still be restored to more than 50\%.

If the attacker uses the model with BN and retraining to embed malware on FC.1 and wants to keep an accuracy loss within 1\% of the model, more than 3150 neurons can be replaced. It will result in $3150\times12/1024=36.9$MB of malware embedded. If the attacker wants to keep the accuracy above 90\%, then 3300 neurons can be replaced, embed $38.7$MB of malware.

\subsection{Security Scan on VirusTotal} We uploaded some of the malware-embedded models to VirusTotal to check whether the malware can be detected. The models were recognized as zip files by VirusTotal. 58 anti-virus engines were involved in the detection works, and no suspicious was detected. It means that this method can evade the security scan by common anti-virus engines.

\begin{table*}
\centering
\caption{Testing Accuracy on different malware}
\label{tab:cmp}
\begin{tabular}{|c|c|c|c|c|c|c|c|c|c|c|}
\hline
 & Method & Model & Base & \begin{tabular}[c]{@{}c@{}}EquationDrug\\ 372KB\end{tabular} & \begin{tabular}[c]{@{}c@{}}ZeusVM\\ 405KB\end{tabular} & \begin{tabular}[c]{@{}c@{}}NSIS\\ 1.7MB\end{tabular} & \begin{tabular}[c]{@{}c@{}}Mamba\\ 2.3MB\end{tabular} & \begin{tabular}[c]{@{}c@{}}WannaCry\\ 3.4MB\end{tabular} & \begin{tabular}[c]{@{}c@{}}VikingHorde\\ 7.1MB\end{tabular} & \begin{tabular}[c]{@{}c@{}}Artemis\\ 12.8MB\end{tabular} \\ \hline
\multirow{8}{*}{EvilModel} & \multirow{8}{*}{\begin{tabular}[c]{@{}c@{}}Fast\\ Substitution\end{tabular}} & Vgg19 & 74.2\% & 74.2\% & 74.2\% & 74.3\% & 74.2\% & 74.2\% & 74.2\% & 74.2\% \\  
 &  & Vgg16 & 73.4\% & 73.4\% & 73.4\% & 73.4\% & 73.4\% & 73.4\% & 73.4\% & 73.4\% \\  
 &  & Alexnet & 56.5\% & 56.5\% & 56.5\% & 56.5\% & 56.5\% & 56.4\% & 56.4\% & 56.4\% \\  
 &  & Resnet101 & 77.4\% & 77.3\% & 77.3\% & 77.2\% & 77.1\% & 77.0\% & 76.7\% & 74.5\% \\ \cline{3-11} 
 &  & Inception & 69.9\% & 69.9\% & 69.9\% & 69.5\% & 69.6\% & 69.1\% & 64.7\% & 61.3\% \\  
 &  & Resnet18 & 69.8\% & 69.7\% & 69.6\% & 69.2\% & 69.1\% & 68.9\% & 67.4\% & 60.3\% \\  
 &  & Mobilenet & 71.9\% & 71.0\% & 71.1\% & 68.5\% & \textbf{60.6\%} & \textbf{39.7\%} & \textbf{0.1\%} & - \\ \hline
\multirow{12}{*}{StegoNet} & \multirow{3}{*}{\begin{tabular}[c]{@{}c@{}}LSB\\ Substitution\end{tabular}} & Inception & 78.0\% & 78.2\% & 77.9\% & 78.0\% & 78.3\% & 78.2\% & 78.1\% & 77.3\% \\  
 &  & Resnet18 & 70.7\% & 69.3\% & 71.2\% & 70.5\% & 72.1\% & 71.3\% & 69.3\% & 61.3\% \\  
 &  & Mobilenet & 70.9\% & \textbf{0.2\%} & \textbf{0.2\%} & \textbf{0.2\%} & \textbf{0.2\%} & \textbf{0.1\%} & - & - \\ \cline{2-11} 
 & \multirow{3}{*}{\begin{tabular}[c]{@{}c@{}}Resilience\\ Training\end{tabular}} & Inception & 78.0\% & 78.3\% & 78.4\% & 78.4\% & 77.6\% & 78.4\% & 77.8\% & 78.1\% \\  
 &  & Resnet18 & 70.7\% & 71.1\% & 71.2\% & 70.4\% & 70.9\% & 71.3\% & 68.2\% & 69.7\% \\  
 &  & Mobilenet & 70.9\% & 71.2\% & 68.5\% & \textbf{32.5\%} & \textbf{6.1\%} & \textbf{0.7\%} & - & - \\ \cline{2-11} 
 & \multirow{3}{*}{\begin{tabular}[c]{@{}c@{}}Value-\\ Mapping\end{tabular}} & Inception & 78.0\% & 78.3\% & 78.4\% & 77.2\% & 78.4\% & 78.1\% & 77.6\% & 77.3\% \\  
 &  & Resnet18 & 70.7\% & 71.1\% & 70.2\% & 72.1\% & 71.0\% & 70.4\% & 70.3\% & 70.9\% \\  
 &  & Mobilenet & 70.9\% & 69.2\% & 71.0\% & \textbf{54.7\%} & \textbf{49.3\%} & - & - & - \\ \cline{2-11} 
 & \multirow{3}{*}{\begin{tabular}[c]{@{}c@{}}Sign-\\ Mapping\end{tabular}} & Inception & 78.0\% & 77.4\% & 78.2\% & 78.0\% & - & - & - & - \\  
 &  & Resnet18 & 70.7\% & 71.1\% & 70.8\% & 69.5\% & - & - & - & - \\  
 &  & Mobilenet & 70.9\% & 68.3\% & - & - & - & - & - & - \\ \hline
\end{tabular}
\end{table*}

\subsection{Comparing with Existing Methods}
Fast substitution has a higher embedding capacity than previous methods. We followed the experiment in StegoNet \cite{Liu20stegonet} to make the comparasion.

\subsubsection{Preparasion} We used 10 pre-trained models (Table~\ref{tab:models}) from public repositories on ImageNet to embed 19 malware samples from Malware DB, and got 180 malware-embedded models. We selected some models randomly and extracted the malware samples from the models. Then we checked the integerity of the malware samples to ensure they have not changed. We used ImageNet to test the performance of the malware-embedded models \textbf{without} retraining them.

Due to the complexity of the models, there is no significant accuracy drop for large-sized models. For medium- and small-sized models, the increased malware in the models has a greater impact on the testing accuracy. However, the embedding capacity is higher than that of StegoNet.

\subsubsection{Comparaing} We selected the commonly used models and malware samples in StegoNet and our work to compare the embedding capacity. As the large models have better fault tolerance, we mainly compare the medium- and small-sized models in this part. The results are shown in Table~\ref{tab:cmp}. The dash-line indicates the model is incapable of embedding the malware due to the size limitation. The bold numbers represent a significant accuracy drop from the baseline. It shows that our method has a higher embedding rate than LSB substitution, value-mapping and sign-mapping. For resilience training, although the embedding rate is at the same level, the testing accuracy of fast substitution is higher than it. Note that the retraining has not been applied to the models, there is still room for improvement in accuracy for fast substitution.

\section{Possible Countermeasures}\label{sec:discuss}
Although malware can be embedded in DNN models, there are still ways to resist such attacks. Firstly, the malware-embedded model cannot be modified. For professionals, the parameters of neurons can be changed through fine-tuning, pruning, model compression or other operations, thereby breaking the malware structure and preventing the malware from recovering normally. However, non-professional users may still face such attacks. Second, the delivery of the malware-embedded models requires methods like supply chain pollution. If users download the required models from a trusted platform and check the model's integrity, the impact can also be reduced. Finally, the DNN model market also needs to verify user identity to avoid being abused by malicious users.

\section{Conclusion}\label{sec:conclusion}

This paper proposes a method that can deliver the malware covertly and evasively through neural network models. When neurons are replaced by malware bytes, the structure of the model remains unchanged. As the malware is disassembled in the neurons, its characteristics are no longer available, which can evade detection by common anti-virus engines. Since the neural network model is robust to changes, there is no significant loss in performance. Experiments show that by applying batch normalization on the fully connected layer, a 178MB-AlexNet model can embed 36.9MB of malware within 1\% accuracy loss. The security scan of VirusTotal also proved that the malware-embedded model can evade detection.

This paper proves that neural networks can also be used maliciously. With the popularity of AI, AI-assisted attacks will emerge and bring new challenges for computer security. Network attack and defense are interdependent. We believe countermeasures against AI-assisted attacks will be applied in the future. We hope the proposed scenario will contribute to future protection efforts.

\bibliographystyle{IEEEtran}
\bibliography{sample-base}

\end{document}